\documentclass[twocolumn]{article}
\usepackage[utf8]{inputenc}
\usepackage{graphicx}
\usepackage{amsmath}
\usepackage{fancyhdr}
\usepackage{float}
\usepackage{stfloats}
\usepackage{booktabs}
\usepackage{amssymb}
\usepackage{tabularx}
\usepackage{listings}
\usepackage{color}
\usepackage[page]{appendix}
\usepackage{subcaption}
\usepackage{natbib}
\usepackage{amsmath}
\usepackage{relsize}
\usepackage[pagebackref=true,breaklinks=true,colorlinks,bookmarks=false]{hyperref}
\newcommand{\pipe}{\mathbin{\vert}}

\definecolor{dkgreen}{rgb}{0,0.6,0}
\definecolor{gray}{rgb}{0.5,0.5,0.5}
\definecolor{mauve}{rgb}{0.58,0,0.82}

\usepackage[
    top=1in,
    bottom=1in,
    left=1in,
    right=1in,
    columnsep=0.2in
]{geometry}

\fancypagestyle{plain}{ 
    \fancyhf{}
    
    \fancyfoot[L]{\textit{Preprint. Under review.} \\ \tt\small{https://deputydev.ai}}
    \fancyfoot[R]{\thepage}
}

\title{\textbf{DeputyDev - AI Powered Developer Assistant:\\
Breaking the Code Review Logjam through Contextual AI to Boost Developer Productivity}}

\author{Vishal Khare, Vijay Saini, Deepak Sharma, Anand Raj, Ankit Rana, Anshul Yadav\\
TATA 1mg Healthcare Solutions Private Limited\\
{\tt\small \{vishal.khare, vijay.saini, deepak.sharma1, anand.raj, ankit.rana1, anshul.yadav\}@1mg.com}
}
\date{}

\begin{document}

\maketitle

\begin{abstract}
This study investigates the implementation and efficacy of DeputyDev, an AI-powered code review assistant developed to address inefficiencies in the software development process. The process of code review is highly inefficient for several reasons like it being a time consuming process, inconsistent feedback and review quality not being at par most of the time. Using our telemetry data we observed that at TATA 1mg, pull request (PR) processing exhibits significant inefficiencies, with average pick-up and review times of 73 and 82 hours respectively, resulting in a 6.2 days closure cycle. These delays negatively impact developer productivity and wellbeing, with context switching during reviews causing substantial time losses. Research from the University of California, Irvine indicates that interruptions can lead to an average of 23 minutes of lost focus  \cite{10.1145/1357054.1357072}, critically affecting code quality and timely delivery.

To address these challenges, we developed DeputyDev's PR review capabilities by providing automated, contextual code reviews. We conducted a rigorous double-controlled A/B experiment involving over 200 engineers to evaluate DeputyDev's impact on review times. The results demonstrated a statistically significant reduction in both per-PR and per-line-of-code review durations. After implementing safeguards to exclude outliers, DeputyDev has been effectively rolled out across the entire organization. Additionally, it has been made available to external companies as a Software-as-a-Service (SaaS) solution, currently supporting the daily work of numerous engineering professionals.

This study explores the implementation and effectiveness of AI-assisted code reviews in improving development workflow timelines and code quality while potentially reducing associated costs.
\end{abstract}

\section{Keywords}
Code review, Contextual AI, Efficiency, Review duration, Code review as SaaS, Developer experience, Delevoper productivity

\section{Introduction}
The software development lifecycle encompasses a multitude of stages, ranging from initial requirement gathering to final production deployment. This process involves numerous critical steps, including code writing, review, and testing. Among these, DeputyDev currently concentrates on enhancing the efficiency of the code review phase.

Code review, while crucial, presents a significant challenge in terms of time consumption, particularly for senior developers \cite{Vijayvergiya_2024}. While there is no one-size-fits-all answer regarding the amount of time spent on code reviews, it is clear that developers typically dedicate a significant portion of their workday, approximately 41 minutes \cite{softwareCodeTime} to this critical activity. The key is balancing thoroughness with efficiency to maintain productivity without compromising code quality. Consequently, code reviews frequently become a bottleneck, impeding rapid releases and continuous integration practices.

A comprehensive code review typically addresses multiple functional and non-functional aspects:

\begin{enumerate}
    \item \textbf{Functional code review}: In functional code review, the developers review the quality of source code by checking whether the syntax of the code is good or not and whether the code follows basic code formatting practices. Functional code reviews are related to the functionality of the software or an application, so if the developer overlooks these code issues, then it might affect the applications.

    \item \textbf{Non-Functional Code Review}: While reviewing the code, the developers analyze the non-functional requirements, such as determining whether the code is secure, scalable, reusable, and easy to maintain. Therefore, the developer strives to ensure that the code is secure, easy to maintain, offers robust performance, and adheres to the best practices laid down in Software Engineering.

\end{enumerate}

Recent advancements in machine learning, particularly in Large Language Models (LLMs), have shown promise in automating code reviews (e.g., studies by Tufano et al.,  \cite{tufano2024codereviewautomationstrengths},  \cite{tufano2022usingpretrainedmodelsboost}; Hong et al., \cite{10.1145/3540250.3549119}). However, the software engineering challenges associated with deploying such systems at scale remain largely unexplored. Furthermore, there is a notable lack of extrinsic evaluations examining the overall effectiveness and user acceptance of these automated systems.

This paper delves into the implementation specifics of DeputyDev's automated contextual code review capabilities \ref{app:capabilities} within the TATA 1mg technology ecosystem. We explore its impact on reducing code review time and provide an in-depth analysis of areas where DeputyDev's performance is suboptimal, along with the underlying reasons for these limitations.

By addressing these aspects, our research aims to bridge the gap between theoretical potential and practical application of LLM-based code review automation, offering valuable insights into its real-world efficacy and challenges.

\section{Hypothesis}
We propose a two-stage code review process that integrates DeputyDev with human expertise. This approach is hypothesized to enhance the efficiency and effectiveness of the overall code review process. The primary mechanisms through which this improvement is expected to occur are:

\begin{enumerate}
    \item \textbf{Reduction of cognitive load on human reviewers}: By employing DeputyDev as a first-level reviewer, we anticipate a significant decrease in the mental effort required from human reviewers. The AI system is expected to handle routine checks and common issues, allowing human reviewers to focus their cognitive resources on more complex, nuanced aspects of the code.
    
    \item \textbf{Minimization of context switching for code authors}: The immediate feedback provided by DeputyDev is hypothesized to reduce the frequency and duration of context switches for code authors. Instead of waiting for human reviewers to provide feedback on routine issues, authors can receive instant guidance on common errors and style violations, allowing them to make necessary adjustments promptly and maintain their coding flow.
    
    \item \textbf{Improved code quality}: The combination of AI and human expertise is expected to result in more thorough and comprehensive code reviews which can result in consistent feedback. DeputyDev can consistently check for a wide range of potential issues, while human reviewers can apply their experience and judgment to more subtle or context-dependent aspects of the code.
\end{enumerate}

\section{Problem}
Code review processes face significant challenges in large-scale, distributed software systems. Organizations with extensive microservice architectures often find that even minor feature requests can trigger changes across multiple services. This complexity makes it impractical to outsource code reviews to external parties, as they would lack the necessary contextual understanding to effectively evaluate the changes.

Consider DeputyDev as this external party. However, DeputyDev orchestrates the creation of optimized context which includes all information required to review the PR.

The importance of context in code review cannot be overstated. It is crucial for comprehending both the immediate changes and their potential ripple effects throughout the entire codebase. Context, in this case, comprises several key pieces of information. We have identified the following items that contribute to making an optimized context for PR review.

\begin{enumerate}
    \item \textbf{PR title}: Title of the pull request articulated by author.
    \item \textbf{PR description}: Description of the pull request articulated by author
    \item \textbf{PR diff}: Output of git diff command between source and destination branches of pull request.
    \item \textbf{Story description}: DeputyDev currently integrates with Jira to pull contents of associated jira stories in order to figure out what needs to be done as per requirements.
    \item \textbf{Approach description}: DeputyDev currently integrates with Confluence to pull contents of associated confluences pages to figure out more details or approach through which change has to be done.
    \item \textbf{Contextually relevant code chunks}: In order to be context-aware, DeputyDev includes contextually relevant code chunks in optimized context.
\end{enumerate}

\begin{figure*}[htbp]
    \centering
    \includegraphics[width=0.8\textwidth]
    {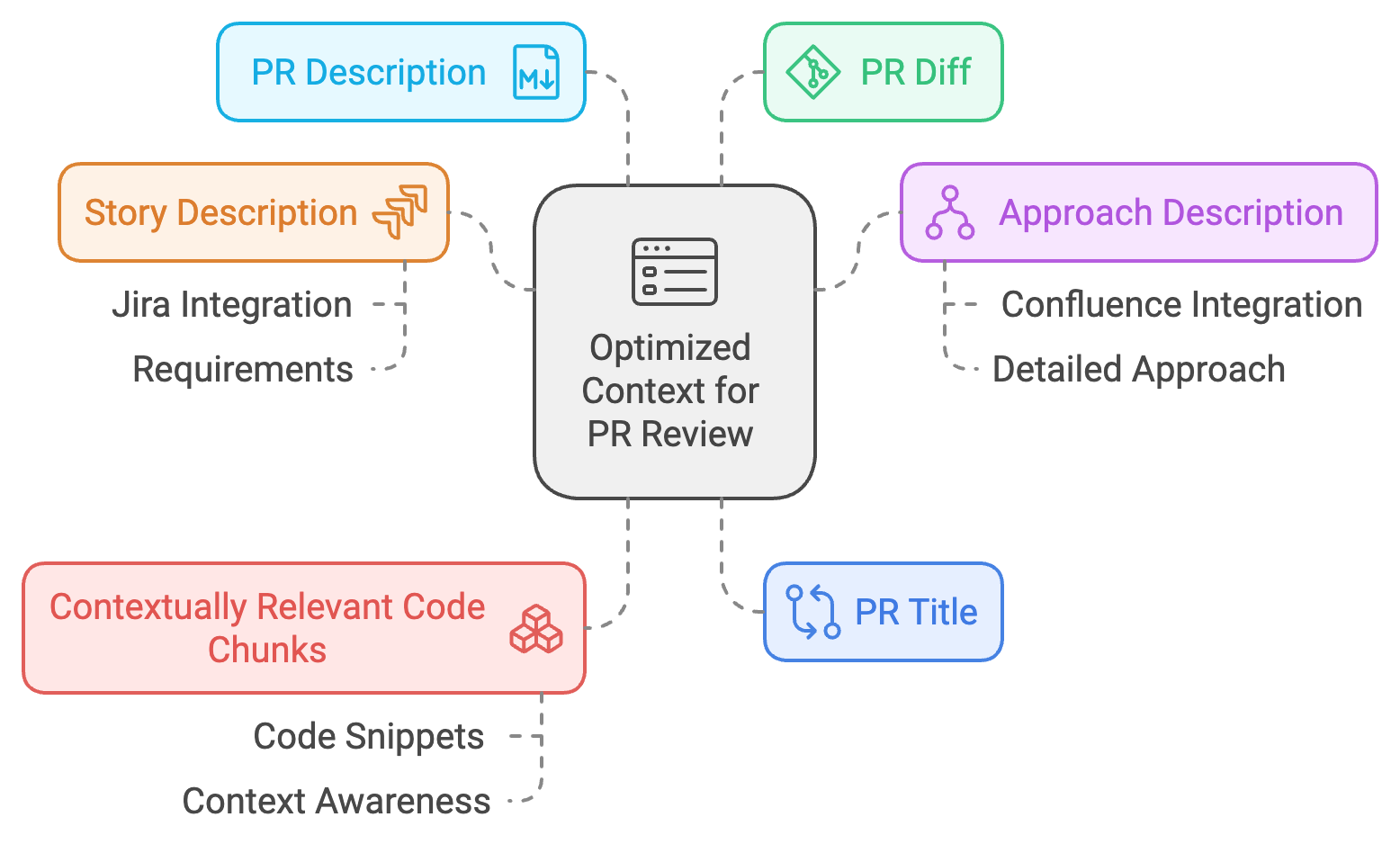}
    \caption{DeputyDev fetches relevant information from multiple sources and unionize them smartly to create an optimized context. This context is self sufficient to provide all necessary information while reviewing PR.}
    \label{fig:context-creation}
\end{figure*}

\section{Approach}
Collating all pieces of context requires extensive integration with sources. For PR title, description and diff, DeputyDev can integrate with popular version control systems like Github, Gitlab and Bitbucket.

For story and approach description DeputyDev integrates with Atlassian suite i.e Jira and Confluence. Every PR being raised should have an associated Jira story or confluence page (mentioned in description of PR). DeputyDev will then pull their contents and format them to be included in optimized context.

Lastly, the most crucial piece of information that makes DeputyDev powerful and context-aware is contextually relevant code chunks. These relevant code chunks are fetched and stored ephemerally while the review process is ongoing. Let us understand with an example what exactly is a contextually relevant code chunk.\ref{fig:context-creation}

Original code
\begin{lstlisting}
def calculate_total(items):
    return sum(item.price for item in items)

def apply_discount(total, discount_percent):
    return total * (1 - discount_percent / 100)

def process_order(items, discount_percent=0):
    total = calculate_total(items)
    final_price = apply_discount(total, discount_percent)
    return final_price

# Example of where process_order is called
class OrderService:
    def create_order(self, user_id, items):
        user = get_user(user_id)
        discount = user.loyalty_discount
        total = process_order(items, discount)
        return Order(user=user, items=items, total=total)
\end{lstlisting}

Diff (Changes to be Made)

\begin{lstlisting}[escapechar=|]
def process_order(items, discount_percent=0):
    total = calculate_total(items)
    if discount_percent > 50:
        raise ValueError("Discount cannot exceed 50%")
    final_price = apply_discount(total, discount_percent)
    return final_price
\end{lstlisting}

Consider a simple code example above and let's break down the contextually relevant code chunks:

\begin{enumerate}
    \item \texttt{calculate\_total} and \texttt{apply\_discount} functions: As mentioned before, these are called within \texttt{process\_order} and remain relevant.
    \item \texttt{OrderService.create\_order} method: This is now a crucial contextually relevant code chunk. It calls \texttt{process\_order} and would be affected by the PR changes in the following ways:
    \begin{itemize}
        \item It might now raise a ValueError if a user's loyalty discount exceeds 50\%.
        \item The error handling in this method (not shown) might need to be updated to catch and handle the new potential ValueError.
    \end{itemize}
    \item \texttt{User} model or class: Although not fully shown, the existence of \texttt{user.loyalty\_discount} suggests there's a User model or class. This becomes contextually relevant because:
    \begin{itemize}
        \item We need to ensure that loyalty discounts in the system never exceed 50\%, or handle cases where they might.
        \item There might be a need to update the User model or associated business logic to cap loyalty discounts at 50\%.
    \end{itemize}
    \item \texttt{Order} model or class: The \texttt{Order} being created with the total from \texttt{process\_order} is relevant because:
    \begin{itemize}
        \item It assumes \texttt{process\_order} always returns a valid total.
        \item If \texttt{process\_order} now throws an exception, we need to ensure the Order creation process can handle this.
    \end{itemize}
\end{enumerate}

This example illustrates how a seemingly small change to \texttt{process\_order} has ripple effects throughout the system. When reviewing the PR, a developer would need to consider:

\begin{itemize}
    \item How to handle potential ValueErrors in the \texttt{OrderService} and potentially other services.
    \item Whether the \texttt{User} model and associated business logic need updates to prevent invalid discount values.
    \item If there are other parts of the system that rely on discounts never being rejected, which might now break.
\end{itemize}

By identifying these contextually relevant code chunks, DeputyDev can perform a more comprehensive review, anticipating potential issues and ensuring that the PR's changes integrate smoothly with the entire system.

A counterpoint to our method of selecting relevant code snippets might be: "Why not just input the entire codebase?" However, this approach faces several challenges:

\begin{enumerate}
    \item \textbf{Limited context window of LLMs}: Though most capable LLMs now support more than 100k context window, It can still prove to be insufficient for large codebases
    \item \textbf{Lost in the middle problem}: The "lost in the middle" problem occurs when LLMs process long texts. They tend to focus on the beginning and end, while losing track of important information in the middle. This is similar to how humans might remember the start and conclusion of a long lecture, but struggle to recall specific details from the middle portion.
    \item \textbf{Cost considerations}: Sending an entire codebase to LLMs as context can be costly.
\end{enumerate}

\begin{figure*}[htbp]
    \centering
    \includegraphics[width=0.8\textwidth]
    {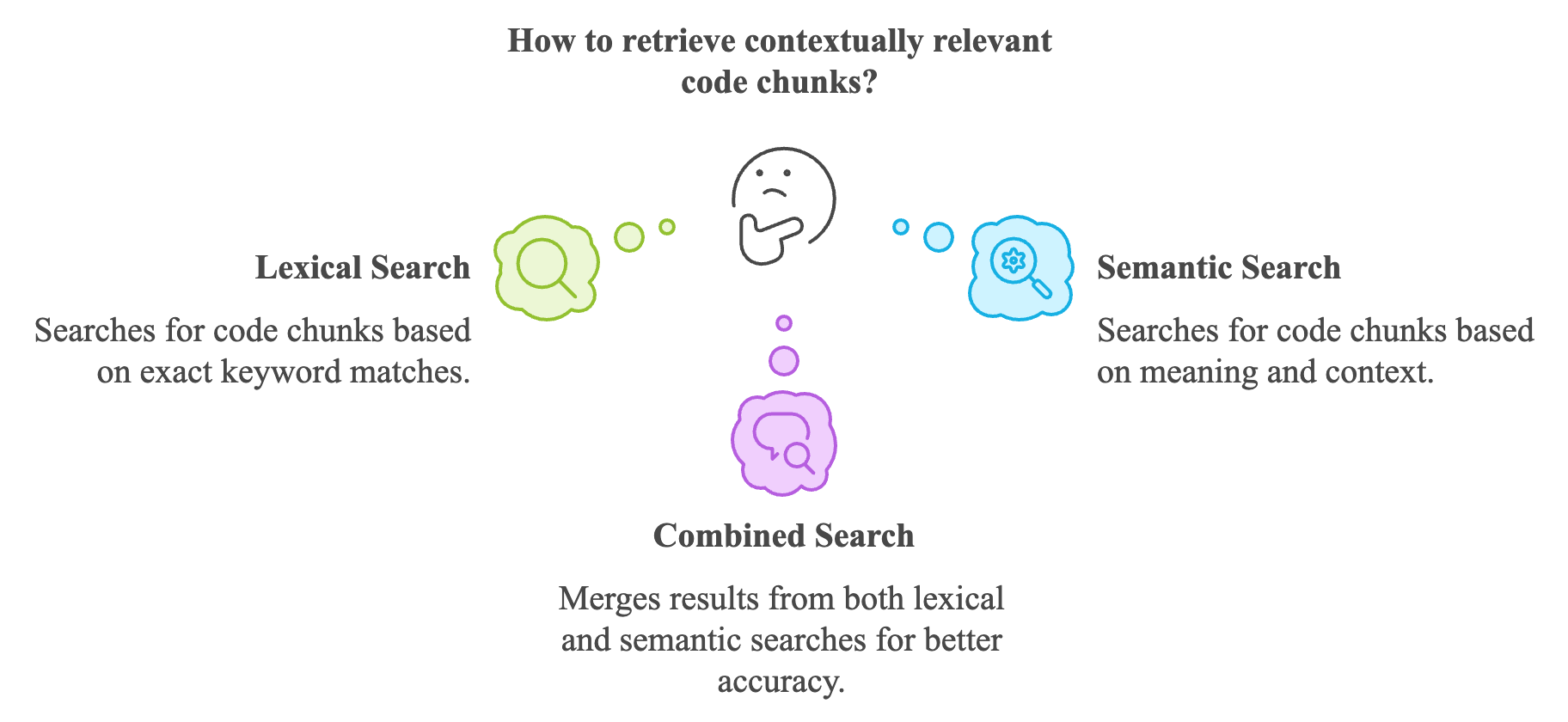}
    \caption{Obtaining contextually relevant code chunks is crucial for code review to happen effectively. This process involves lexical and semantic search and then combining both their results together.
}
    \label{fig:DeputyDev-search}
\end{figure*}

\section{Implementation Details}
As mentioned earlier, DeputyDev can integrate with popular version control systems like Github, Gitlab and Bitbucket. When a pull request is raised, DeputyDev receives an event through webhook. Automated review then starts in the background without interfering with the developer's usual workflow.

Upon receiving this pull request event, DeputyDev starts to create optimized context as mentioned above by leveraging its integration with version control systems and Atlassian suite

\subsection{Retrieving contextually relevant code chunks}
Everytime a pull request event is received by DeputyDev, It clones the repository ephemerally, create Abstract Syntax Trees(AST) out of it and semantically chunk the ASTs. This process of breaking down the code into meaningful, logical segments based on their functionality, purpose, or conceptual relationships, rather than just arbitrary divisions. Each chunk is provided with a unique ID and is stored in the redis cache with a specified TTL.

After performing chunking of the codebase several types of searches including lexical and semantic search is performed on the set of chunks. These searches is performed by using PR diff as a search query. Search results then received are merged together to obtain contextually relevant code chunks.

Example - If $R_{LS}$ is set of code chunks retrieved by lexical search and $R_{SS}$ is set of code chunks retrieved by semantic search then -

$R_f = (R_{LS} \cup R_{SS})$

Where, $R_f$ is final set of contextually relevant code chunks\ref{fig:DeputyDev-search}

Now that we have all the pieces of information we need to create optimized context, the next step would be to call LLMs with this context to get a response.

\subsection{Agentic workflow}
DeputyDev uses agentic workflow to further optimize the quality and better segregate responsibilities of different aspects of the code review process. Inspired by Andrew Ng's letters \cite{deeplearningFourAgent}, DeputyDev's agentic workflow is an amalgamation of 2 agentic design patterns i.e. Multi-agent and Reflection.

An agentic workflow is different from a comprehensive workflow where optimized context is passed on to LLM to obtain review on all aspects of code in a single pass. In an agentic workflow, we define agents to cater to each specific aspect. There are pros and cons to agentic design which are as follows-

\subsubsection{Benefits of Agentic Workflow}
\begin{enumerate}
    \item \textbf{Attention and Working Memory}: AI models, like humans, have limitations in their "working memory" (in this case, the context window). By focusing on one aspect at a time, the model can dedicate more of its processing capacity to that specific task, potentially leading to more thorough and nuanced analysis.
    \item \textbf{Task Complexity}: Code review involves multiple complex subtasks. Breaking these down into separate focused reviews aligns with the principle of divide-and-conquer in problem-solving, which often leads to better results in complex tasks.
    \item \textbf{Prompt Engineering}: This method allows for more specific and detailed prompts for each aspect, which can guide the model's attention more effectively. This is similar to how specific questions often elicit more precise and useful responses in human communication.
    \item \textbf{Iterative Improvement}: This method allows for iterative refinement of prompts based on the results of each focused review, potentially leading to better overall outcomes.
    \item \textbf{Parallel Processing}: Agents can work simultaneously on different aspects of the code, potentially reducing overall review time.
    \item \textbf{Customizable Reviews}: Users can choose which agents to activate based on their specific needs or project requirements.
    \item \textbf{Clearer Reporting}: Issues can be categorized more effectively, making it easier for developers to understand and address feedback.
    \item \textbf{Continuous Improvement}: Individual agents can be updated or fine-tuned without affecting the entire system.
\end{enumerate}

\subsubsection{Drawbacks of Agentic Workflow}
\begin{enumerate}
    \item \textbf{Increased Complexity}: Managing multiple agents adds complexity to the system architecture and maintenance.
    \item \textbf{Increased Inference cost}: Multiple agents means all of them will have their own optimized context. This will surely increase the number of input tokens and may increase the number of output tokens.
    \item \textbf{Potential Redundancy}: Agents might overlap in their analyses, leading to duplicate findings or conflicting advice.
    \item \textbf{Integration Challenges}: Ensuring smooth communication and coordination between agents may be challenging.
    \item \textbf{Higher Resource Usage}: Running multiple agents may require more computational resources compared to a single, comprehensive review.
\end{enumerate}

After carefully considering the advantages and disadvantages of an agentic workflow, we decided to upgrade to a generic design approach from a comprehensive approach as it is superior for achieving optimal quality. Consequently, we decided to implement an agentic workflow strategy.

Let's discuss agentic design patterns we used-

\begin{figure*}[htbp]
    \centering
    \includegraphics[width=0.8\textwidth]
    {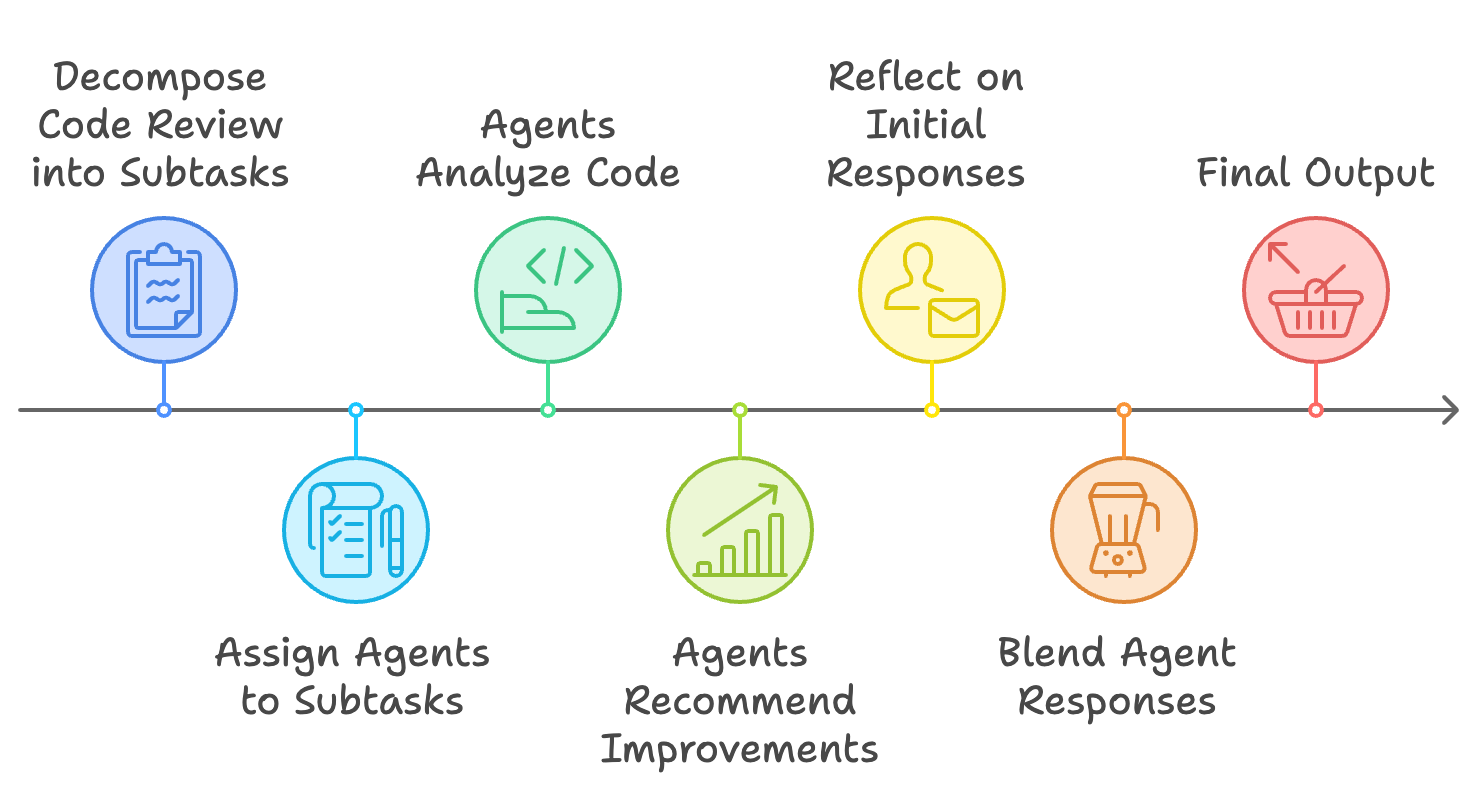}
    \caption{DeputyDev workflow using multi-agent and reflection design patterns before blending the results together.}
    \label{fig:DeputyDev-workflow}
\end{figure*}

\subsubsection{Multi-Agent design pattern}
\textit{"Even though some LLMs today can accept very long input contexts (for instance, Gemini 1.5 Pro accepts 1 million tokens), their ability to truly understand long, complex inputs is mixed. A multi-agentic workflow in which the LLM is prompted to focus on one thing at a time can give better performance. By telling it when it should play software engineer, we can also specify what is important in that role's subtask. For example, the prompt above emphasized clear, efficient code as opposed to, say, scalable and highly secure code. By decomposing the overall task into subtasks, we can optimize the subtasks better."} - Andrew Ng \cite{deeplearningAgenticDesign}

We decompose our task of reviewing code into smaller subtasks. Each subtask being handled by an agent of its own. \cite{qian2024chatdevcommunicativeagentssoftware} Following are the agents we defined-

\begin{enumerate}
    \item \textbf{Security}: This agent is responsible to identify and recommend corrective code for any security issues it observes in the code. This could be issues like Injection attacks, Improper input validation, Using components with known vulnerabilities, Hardcoded credentials etc.
    \item \textbf{Code communication}: This agent is responsible to identify issues with aspects like documentation, docstring and logging. This could be issues like missing/incomplete docstrings, Improper documentation for a snippet. Incorrect or inconsistent logging usage etc.
    \item \textbf{Performance optimizations}: This agent is responsible to identify issues specifically related to performance issues and suggest optimisations on them. This will include algorithmic optimisations to reduce space \& time complexity, database query optimizations and other forms of code optimisation best practices.
    \item \textbf{Code maintainability}: This agent is responsible to identify issues pertaining to code maintainability, readability, reusability and quality.
    \item \textbf{Errors}: This agent is responsible to identify any logical, syntactical or runtime error that the change may induce.
    \item \textbf{Business validation}: This agent is responsible for correlating changes being done in PR to actual business requirements. It checks if changes adhere to stated requirements or not.
\end{enumerate}

\begin{table*}[ht]
\centering
\caption{List of agents along with which piece of information constitutes to make optimized context.}
\label{tab:agent-prompt-matrix}
\small
\begin{tabularx}{\textwidth}{@{}l*{7}{>{\centering\arraybackslash}X}@{}}
\toprule
Agent & PR Diff & PR Title & PR Description & Context Code & User Story & Confluence Pages & Initial LLM Review \\
\hline
Security & $\checkmark$ & $\checkmark$ & $\checkmark$ & & & & $\checkmark$ (Refl.) \\
Code Communication & $\checkmark$ & $\checkmark$ & $\checkmark$ & & & & $\checkmark$ (Refl.) \\
Performance Optimization & $\checkmark$ & $\checkmark$ & $\checkmark$ & $\checkmark$ & & & $\checkmark$ (Refl.) \\
Code Maintainability & $\checkmark$ & $\checkmark$ & $\checkmark$ & $\checkmark$ & & & $\checkmark$ (Refl.) \\
Error & $\checkmark$ & $\checkmark$ & $\checkmark$ & $\checkmark$ & & & $\checkmark$ (Refl.) \\
Business Logic Validation & $\checkmark$ & $\checkmark$ & $\checkmark$ & $\checkmark$ & $\checkmark$ & $\checkmark$ & $\checkmark$ (Refl.) \\
\bottomrule
\end{tabularx}
\end{table*}

\subsubsection{Reflection design pattern}
\textit{"Today, we mostly use LLMs in zero-shot mode, prompting a model to generate final output token by token without revising its work. This is akin to asking someone to compose an essay from start to finish, typing straight through with no backspacing allowed, and expecting a high-quality result. Despite the difficulty, LLMs do amazingly well at this task!"} - Andrew Ng \cite{deeplearningFourAgent}

Reflection in the context of large language models (LLMs) refers to the ability of an AI system to analyze and think about its own thought processes, outputs, and performance. This concept aims to make AI models more self-aware and capable of improving their own responses. \cite{shinn2023reflexionlanguageagentsverbal}  \cite{madaan2023selfrefineiterativerefinementselffeedback} Here are some key aspects of reflection in LLMs:

\begin{enumerate}
    \item \textbf{Self-evaluation}: The model assesses the quality, accuracy, and appropriateness of its own outputs.
    \item \textbf{Iterative improvement}: Based on self-evaluation, the model can refine and improve its responses.
    \item \textbf{Metacognition}: The ability to think about its own thinking process, including recognizing limitations or potential biases.
    \item \textbf{Error detection}: Identifying mistakes or inconsistencies in its own reasoning or outputs.
    \item \textbf{Confidence assessment}: Evaluating how certain the model is about different parts of its response.
\end{enumerate}

Upon receiving the initial response from LLMs, we send it back to LLM with its initial response and ask it to reflect on it. LLM then responds back with a higher quality response. This is done for each agent in order to achieve higher quality output.

GPT-3.5 (zero shot) was 48.1\% correct. GPT-4 (zero shot) does better at 67.0\%. However, the improvement from GPT-3.5 to GPT-4 is dwarfed by incorporating an iterative agent workflow. Indeed, wrapped in an agent loop, GPT-3.5 achieves up to 95.1\% \cite{deeplearningFourAgent}


\subsection{Structured/Unstructured output}
LLMs output are string or stream of strings. To use LLM responses for anything meaningful in industrial grade applications, a structured (JSON/XML) output is required for it to be successfully parsed by code. Developers have long been working around the limitations of LLMs in this area via open source tooling, prompting, and retrying requests repeatedly to ensure that model outputs match the formats needed to interoperate with their systems.

OpenAI and Anthropic models already have a JSON mode hyperparameter which can be turned on to expect a JSON response. While JSON mode improves model reliability for generating valid JSON outputs, it does not guarantee that the model's response will conform to a particular schema. OpenAI recently introduced structured output in its API \cite{Introduc39:online}

It is however noticed that enforcing schema restriction on LLM's response results in significant drop in LLM's reasoning capabilities. \cite{tam2024letspeakfreelystudy} Thus, it is required to request LLMs without any schema restrictions and separate out the process of enforcing schema from the reasoning step.

Refer to Appendix \ref{app:agent_output_schema} for a LLM agent's response when schema restriction is enforced

\subsection{Blending engine}
Blending engine ($\Sigma$) is an integral part of DeputyDev's agentic workflow. It acts as a consolidator of final comments to be made on the pull request. Moreover, It provides us ways and measures by which we can filter in/out and fine tune the final comments.

Upon receiving responses from all the agents, all responses are unionized and passed on to the blending engines. Blending engine has the capability to house dimensions in it. These dimensions are basically a set of rules that each agentic response should follow. As a result of running each dimension, comments can be filtered in/out or modified. This is done to achieve higher quality response and reduce noise in most cases.

Following are some dimensions of blending engine-

\begin{enumerate}
    \item \textbf{Confidence score}: With the help of prompt engineering we are able to instruct LLMs to respond back with a confidence score with each comment. This is basically LLMs self confidence in the comment it is returning. This is a relatively simple dimension where we filter out all comments having a confidence score below some value X. This value X varies between agents depending on its weightage.
    \item \textbf{Comment overlap summarisation}: There can be a case where multiple agents provide their respective comments on the same line of a file. This can add noise for the reviewers and authors of the PR. This dimension summarizes all overlapping comments into one.
\end{enumerate}

\subsection{Mathematical modeling of agentic design}
To structure our agents, we refer communicative agents \cite{qian2024chatdevcommunicativeagentssoftware}

\begin{enumerate}
    \item $R_f$ is the final output (a list of comments that DeputyDev will make on the PR)
    
    $R_f = \Sigma\langle A_1, A_2, ..., A_n\rangle$
    
    \begin{itemize}
        \item Where $A_i$ is an agent responsible for performing review on PR from a specific perspective like security, code quality etc. Agents are mutually exclusive and can be executed in parallel.
        \item $\Sigma$ refers to a blending engine which acts as a post-processing module and consolidates (filter in/out) comments basis some predefined dimensions/factors like confidence score etc.
    \end{itemize}
    
    \item $A_i = C\langle R_{sp}, R_r\rangle$
    \begin{itemize}
        \item Where C is a consensus reached between $R_{sp}$ (Single pass response from LLM) and $R_r$ (Reflection response from LLM)
    \end{itemize}
    
    \item $C\langle R_{sp}, R_r\rangle = \langle DD \rightarrow LLM, LLM ; DD\rangle_{sp} | \langle DD \rightarrow LLM, LLM ; DD\rangle_r$
    \begin{itemize}
        \item Where DD is DeputyDev and LLM can be any large language model. Claude 3.5 Sonnet in our case.
        \item This means DD submits a prompt to LLM $\langle DD \rightarrow LLM\rangle$ and LLM responds back to DD $\langle LLM ; DD\rangle$
        \item Subscripted sp and r means Single pass (1st request to LLM) and reflection response respectively. Pipe $ \pipe $ represents reflection. It means Single pass response is fed back to LLM as context to attain structured and higher quality response.
        \item Reflection in this case serves 2 objectives-
        \begin{itemize}
            \item Self reflect on LLM's response and correct the response against any anomalies. It increases the quality of response.
            \item Converting unstructured response to structured response for DD to consume.
        \end{itemize}
    \end{itemize}
\end{enumerate}

\section{Additional features}

\subsection{Context-aware chatting}
Authors or reviewers of PRs can also chat with DeputyDev by initializing with \#dd or \#deputydev and then their prompt. This is a powerful feature which can not only answer any technical questions while being completely context-aware but can also generate code, tests and documentations or just act as a learning companion for developers.

Some examples are as follows-
\begin{itemize}
    \item \#dd - Why are parameterized queries safer?
    \item \#dd - Generate a docstring for this function.
\end{itemize}

Refer to Appendix \ref{app:capabilities} for screenshots of above examples

\subsection{PR summary}
DeputyDev also generates impeccable PR summaries along with its size w.r.t. LOCs changed in the PR and estimated time to review the PR. This is particularly helpful for reviewers and service owners to get a gist of the change being made quickly at a glance without actually going through the diff.

Refer to Appendix \ref{app:capabilities} for screenshots of PR summary

\section{Choice of model}
DeputyDev currently integrates with OpenAI and Anthropic. We use OpenAI's GPT-4o for PR summarization because of its superior summarization capabilities and Anthropic's claude 3.5 sonnet to perform actual code reviews due to its superior performance on HumanEval benchmark \cite{anthropic}

Both models are highly capable and best in the industry right now. DeputyDev does not use any fine-tuned model as of now.

\section{Experiment setting}
To measure DeputyDev's impact on time taken to review a PR we designed a double controlled A/B experiment. Following are some aspects of experiment settings-

\begin{enumerate}
    \item TATA 1mg uses Atlassian's Bitbucket as its version control system.
    \item 3 sets were defined namely Control 1, Control 2 and Test sets with 33\% allocation to each one of them.
    \item This essentially means that 33\% PRs of a repository will be reviewed by DeputyDev.
    \item The final analysis excluded pull requests (PRs) that were considered outliers based on their size, as measured by the number of lines of code (LOC) changed. Specifically, PRs in the top 25 percentile and bottom 10 percentile of size were removed from the dataset before conducting the final analysis.
    \item We excluded repositories that lacked balanced representation across all three sets. A repository was considered balanced only if each set contained at least 10 or an equal number of pull requests. This filtering process helped us achieve statistical normalization of behavior among the sets.
    \item Data collated is of comprehensive workflow and while using OpenAI's GPT-4o as LLM.
    \item Duration of the experiment is 30 days from 27th July 2024 to 27th August 2024.
\end{enumerate}

\section{Experiment results}

\subsection{Distribution of pull requests by size in sets}
\begin{figure}[htbp]
    \centering
    \includegraphics[width=1\columnwidth]
    {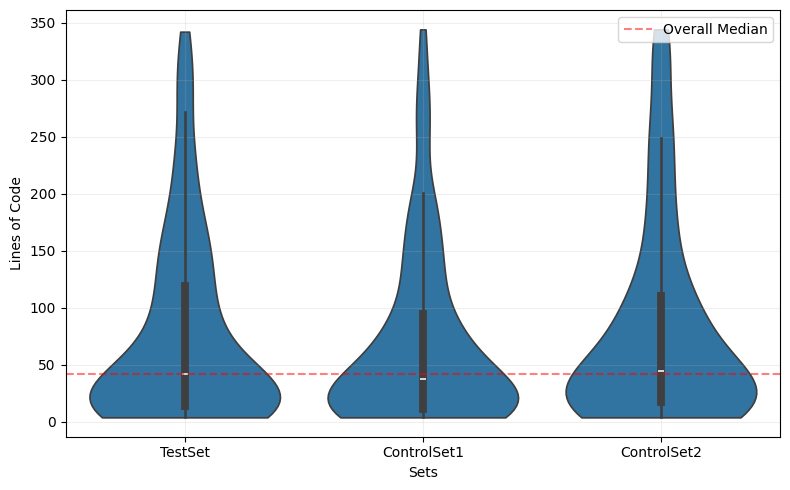}
    \caption{Distribution of PRs by sizes across sets. We can observe all 3 sets are similar}
    \label{fig:pr_dist_across_sets}
\end{figure}

To validate the statistical significance of our experiment, we examined the distribution of pull requests by size across all sets. Our analysis revealed a uniform distribution, lending credence to the robustness of our experimental design and the validity of our findings. This uniformity in size distribution across sets strengthens the reliability of our statistical inferences and the overall significance of the observed results. Refer to Figure \ref{fig:pr_dist_across_sets}.

\begin{figure*}[htbp]
    \centering
    \begin{subfigure}[t]{\columnwidth}
        \centering
        \includegraphics[scale=0.35]{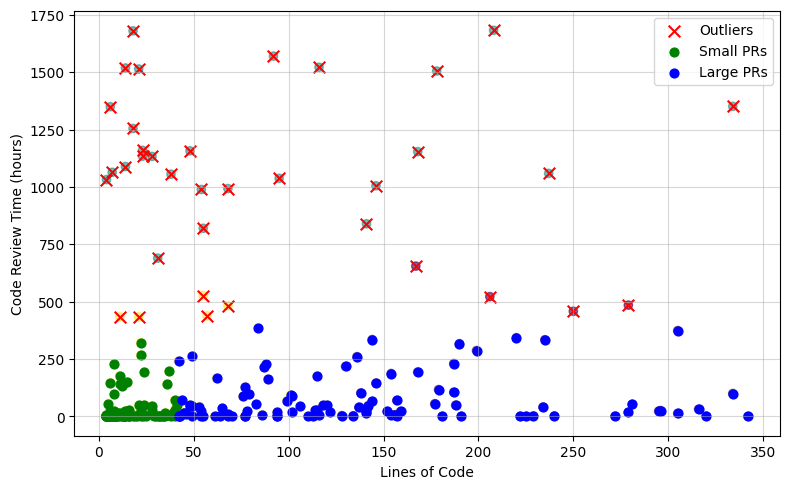}
        \caption{Test set with correlation coefficient of 0.095}
    \end{subfigure}%
    ~ 
    \begin{subfigure}[t]{\columnwidth}
        \centering
        \includegraphics[scale=0.35]{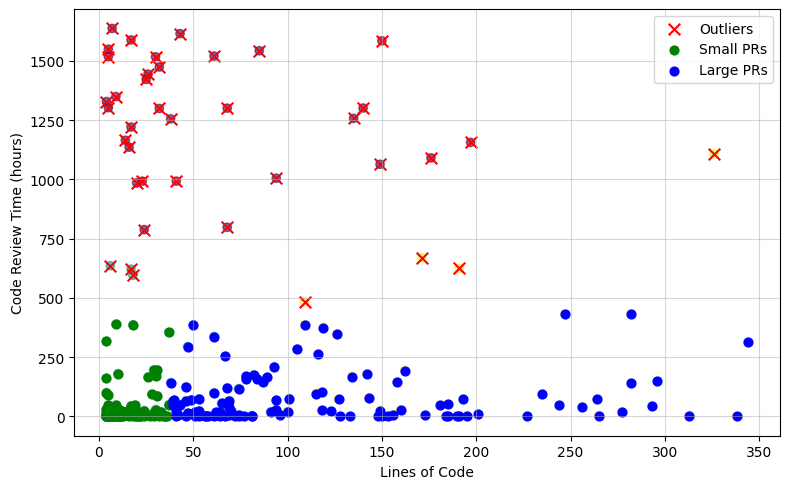}
        \caption{Control 1 set with correlation coefficient of 0.004}
    \end{subfigure}

    \centering
    \begin{subfigure}[t]{\columnwidth}
        \centering
        \includegraphics[scale=0.35]{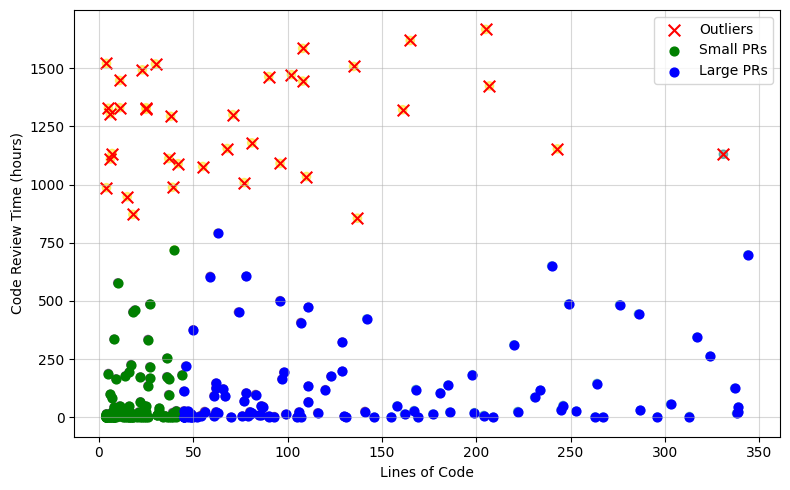}
        \caption{Control 2 set with correlation coefficient of 0.052}
    \end{subfigure}%

    \caption{Weak correlation between lines of code changed in a pull request vs time to review that PR}
\label{fig:loc_vs_review_time_distribution}    
\end{figure*}

\subsection{Weak correlation between lines of code and code review time}
While one might expect a strong positive correlation between the number of lines of code to be reviewed and the time required for review, empirical evidence suggests otherwise. Contrary to intuition, experienced professionals in the field often observe a weak relationship between these variables. Our experimental results corroborate this phenomenon. As illustrated in Figure \ref{fig:loc_vs_review_time_distribution}, the correlation coefficients between lines of code changed and code review time for pull requests are notably low: 0.095, 0.004, and 0.052 for the test, control 1, and control 2 sets, respectively. These findings indicate a minimal association between the volume of code changes and the duration of the review process.

\subsection{Time savings due to DeputyDev intervention}
Our analysis extended to examining the average code review time per pull request (PR), per line of code, and the median code review time. The results revealed a statistically significant reduction in all these temporal metrics, primarily attributable to DeputyDev's interventions. This observation substantiates the hypothesis that DeputyDev's involvement in PR reviews leads to expedited turnaround times. A key factor contributing to this efficiency is the immediate feedback provided to PR authors upon submission, eliminating the delay typically associated with waiting for a human reviewer to initiate the review process. This immediate engagement appears to streamline the overall review workflow. Refer to Table \ref{tab:pr_review_stats} for numbers and Appendix \ref{app:calculations} for calculations.

\begin{table*}[ht]
\centering
\begin{tabularx}{\textwidth}{@{}l*{7}{>{\centering\arraybackslash}X}@{}}
\toprule
\hline
\textbf{Set} & \textbf{No. of PRs} & \textbf{Avg Loc Per PR} & \textbf{Avg Review Time (hrs)} & \textbf{Avg Review Time Per LOC (hrs)} & \textbf{Median Review Time (hrs)} \\
\hline
ControlSet1 (CS1) & 244 & 69 & 239.57 & 12.97 & 0.76 \\
ControlSet2 (CS2) & 238 & 82 & 278.14 & 12.29 & 0.78 \\
TestSet (TS) & 239 & 78 & 197.97 & 7.50 & 0.41 \\
\hline
\textbf{TS $\Delta$ CS1 (\%)} & - & - & -17.36 & -42.19 & -46.36\\
\textbf{TS $\Delta$ CS2 (\%)} & - & - & -28.82 & -38.98 & -47.52\\
\hline
\end{tabularx}
\caption{We observe significant decrease in avg review time, avg review time per LOC and median review time in Test set upto 28.82\%, 42.19\% and 47.52\% respectively}
\label{tab:pr_review_stats}
\end{table*}

\begin{table*}[ht]
\centering
\begin{tabularx}{\textwidth}{@{}l*{6}{>{\centering\arraybackslash}X}@{}}
\toprule
\hline
\textbf{PR size category(LOC)} & 
\textbf{CS1} & 
\textbf{CS2} & 
\textbf{TS} & 
\textbf{TS $\Delta$ CS1 (\%)} & 
\textbf{TS $\Delta$ CS2 (\%)} & 
\textbf{PR Count} \\
\hline
0-50 (S) & 21.23 & 20.34 & 11.92 & -43.87 & -41.40 & 399 \\
\hline
51-100 (M) & 2.61 & 4.09 & 3.50 & 34.01 & -14.57 & 123 \\
\hline
101-200 (L) & 2.01 & 3.00 & 1.44 & -28.42 & -52.08 & 127 \\
\hline
201-500 (XL) & 0.58 & 1.27 & 1.17 & 100.30 & -8.29 & 72 \\
\hline
\end{tabularx}
\caption{Significant dent made by DeputyDev in PR categories S and L while having mixed performance in size category M and XL. It is however evident that DeputyDev is more effective for PRs in category M vs category XL}
\label{table:pr_stats_by_size}
\end{table*}

\subsection{Distribution of pull requests by size}
Another inquiry in our study was to determine the optimal pull request size for DeputyDev's intervention. Our findings indicate that DeputyDev's efficacy is inversely proportional to the size of the pull request. Specifically, we observed that DeputyDev's impact is most pronounced on smaller pull requests, while its effectiveness, though still notable, diminishes for larger ones.

Our analysis reveals that while larger PRs have a more substantial impact on the overall average review time, smaller PRs disproportionately influence the average review time per line of code (LOC). This phenomenon can be attributed to the fixed costs associated with context-switching and initial PR setup, which tend to result in a higher time per LOC ratio for smaller PRs. Refer table \ref{table:pr_stats_by_size}

To illustrate this mathematically, consider a simplified example:

As an example - Prior to improvements, a small PR of 10 LOC might require 10 minutes (1 min/LOC), while a larger PR of 100 LOC could take 50 minutes (0.5 min/LOC). Post-improvement, the same small PR might only require 5 minutes (0.5 min/LOC), representing a 50\% reduction in time per LOC. Conversely, the larger PR might now take 45 minutes (0.45 min/LOC), showing a 10\% reduction in time per LOC.

Calculating the average review times, we observe a decrease from 30 minutes to 25 minutes, representing a 16.7\% reduction. However, when examining the average review time per LOC, we note a more substantial decrease from 0.75 min/LOC to 0.475 min/LOC, equating to a 36.7\% reduction.

This discrepancy between the reduction in average review time (16.7\%) and the reduction in average review time per LOC (36.7\%) can be attributed to several factors:

\begin{itemize}
    \item \textbf{Reduction of fixed costs:} Each pull request (PR) review incurs fixed overhead costs, irrespective of its size, with context-switching being the most significant among these. These fixed costs disproportionately affect smaller PRs, resulting in a higher time-to-lines-of-code (LOC) ratio. The observed reduction in this ratio suggests that DeputyDev has effectively mitigated context-switching to some degree and other associated fixed costs, thereby improving overall review efficiency.
    \item \textbf{Cognitive load management:} Smaller PRs may benefit more significantly from enhancements in review tools or processes that facilitate rapid context comprehension and change assessment.
    \item \textbf{Proportional impact:} Time savings on smaller PRs translate to larger percentage improvements in time per LOC compared to equivalent time savings on larger PRs.
\end{itemize}

These findings suggest that DeputyDev is particularly effective in optimizing the efficiency of smaller PR reviews, while still offering benefits across all PR sizes. Refer table \ref{table:pr_stats_by_size} for experiment numbers.

\section{Conclusion}
This study has identified and addressed significant inefficiencies in the conventional code review process widely adopted across the software industry. These inefficiencies not only result in considerable time losses but also contribute to developer frustration. To mitigate these issues, we have introduced DeputyDev, an innovative automated tool for contextual code reviews triggered by pull request submissions. We have detailed the implementation of DeputyDev, highlighting its use of Abstract Syntax Trees (AST) and semantic search to generate optimized context. Furthermore, we have demonstrated DeputyDev's seamless integration with Atlassian suite products such as Jira and Confluence, enabling validation of code changes against business requirements.

To evaluate the effectiveness of DeputyDev, we conducted a rigorous double-controlled A/B experiment. The results revealed a substantial reduction in the time required for pull request reviews when utilizing DeputyDev. These findings underscore the significant value that DeputyDev can offer to various stakeholders in the development lifecycle, including developers and reviewers. Our research thus presents DeputyDev as a promising solution to enhance efficiency and satisfaction in the code review process, potentially transforming current industry practices.

{\small
\bibliographystyle{plain}
\bibliography{egbib.bib}
}

\clearpage

\onecolumn

\appendix

\begin{appendices}

\section{Output from LLM agents}
\label{app:agent_output_schema}
\lstdefinestyle{XMLStyle}{
  language=XML,
  basicstyle=\ttfamily\small,
  numbers=left,
  numberstyle=\tiny,
  stepnumber=1,
  numbersep=5pt,
  tabsize=2,
  extendedchars=true,
  breaklines=true,
  keywordstyle=\color{blue},
  stringstyle=\color{orange},
  commentstyle=\color{green},
  morestring=[b]",
  showspaces=false,
  showtabs=false,
  showstringspaces=false
}
This document presents the output structure from an LLM (Large Language Model) agent's code review. Each field in the structure has a specific significance:

\begin{itemize}
    \item \textbf{description}: Provides a detailed explanation of the issue or suggestion identified by the LLM agent.
    \item \textbf{corrective\_code}: Contains the proposed code changes or additions to address the identified issue (if any).
    \item \textbf{file\_path}: Indicates the specific file where the issue was found or where changes should be applied.
    \item \textbf{line\_number}: Specifies the exact line in the file where the issue occurs or where changes should be made.
    \item \textbf{confidence\_score}: Represents the LLM agent's confidence level in its assessment or suggestion.
    \item \textbf{bucket}: Categorizes the type of issue or suggestion (e.g., Readability, Performance optimization, Business requirement adherence, Documentation etc.).
\end{itemize}

\section*{LLM Agent Review Output}

\begin{lstlisting}[style=XMLStyle]
<review>
  <comments>
    <comment>
      <description></description>
      <corrective_code></corrective_code>
      <file_path></file_path>
      <line_number></line_number>
      <confidence_score></confidence_score>
      <bucket></bucket>
    </comment>
  </comments>
</review>
\end{lstlisting}

\section{Capabilities}
\label{app:capabilities}
\subsection{Contextual code reviews}
\begin{figure*}[htbp]
    \centering
    \includegraphics[scale=0.40]
    {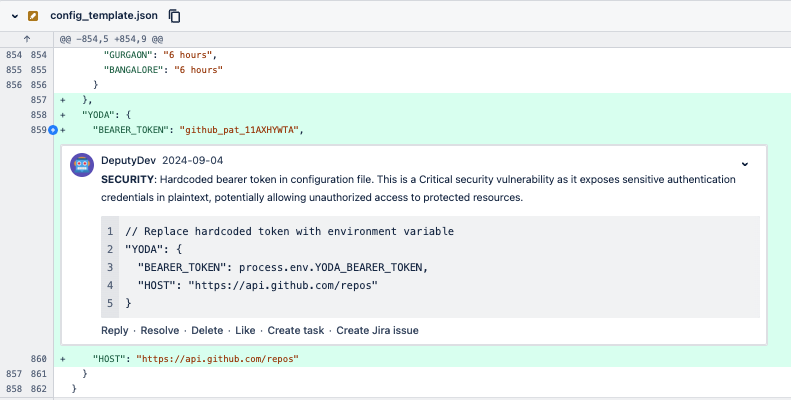}
    \caption{DeputyDev allow users to adhere to secure coding practices}
    \label{fig:dd_comm_security}
\end{figure*}

\begin{figure*}[htbp]
    \centering
    \includegraphics[scale=0.40]
    {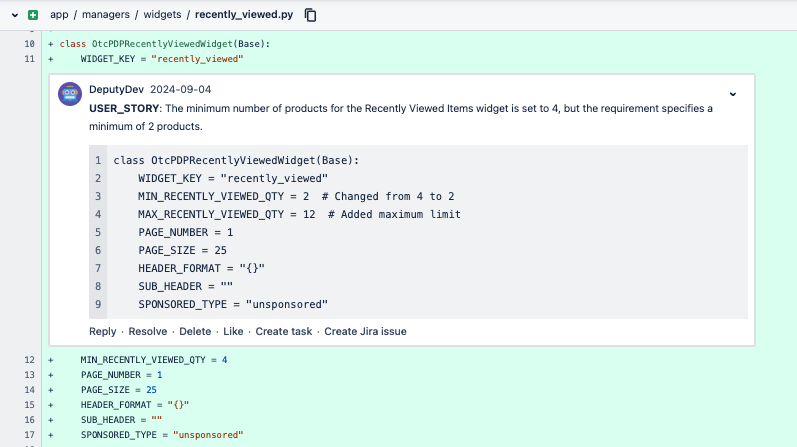}
    \caption{DeputyDev ensures business logic adherence. Making sure changes are done in accordance to task description.}
    \label{fig:dd_comm_story}
\end{figure*}

\begin{figure*}[htbp]
    \centering
    \includegraphics[scale=0.40]
    {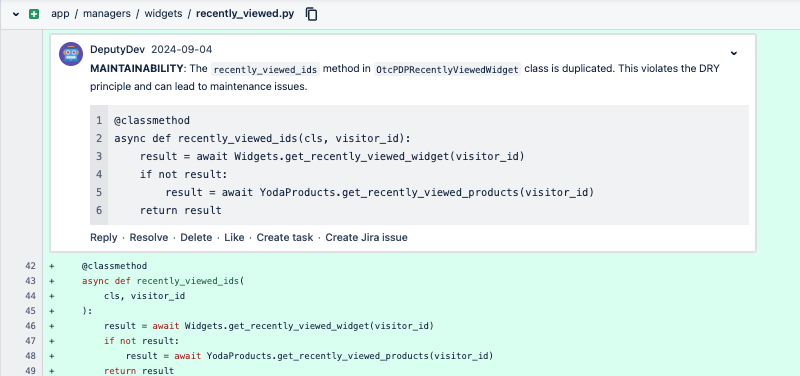}
    \caption{DeputyDev identifying code duplication issue.}
    \label{fig:dd_comm_maint}
\end{figure*}

DeputyDev's core feature is to perform highly contextual code review and provide in-line comments indicating issues or potential improvements. \ref{fig:dd_comm_security} \ref{fig:dd_comm_story} \ref{fig:dd_comm_maint}

\subsection{Context-aware chatting}
\begin{figure*}[htbp]
    \centering
    \includegraphics[width=0.8\textwidth]
    {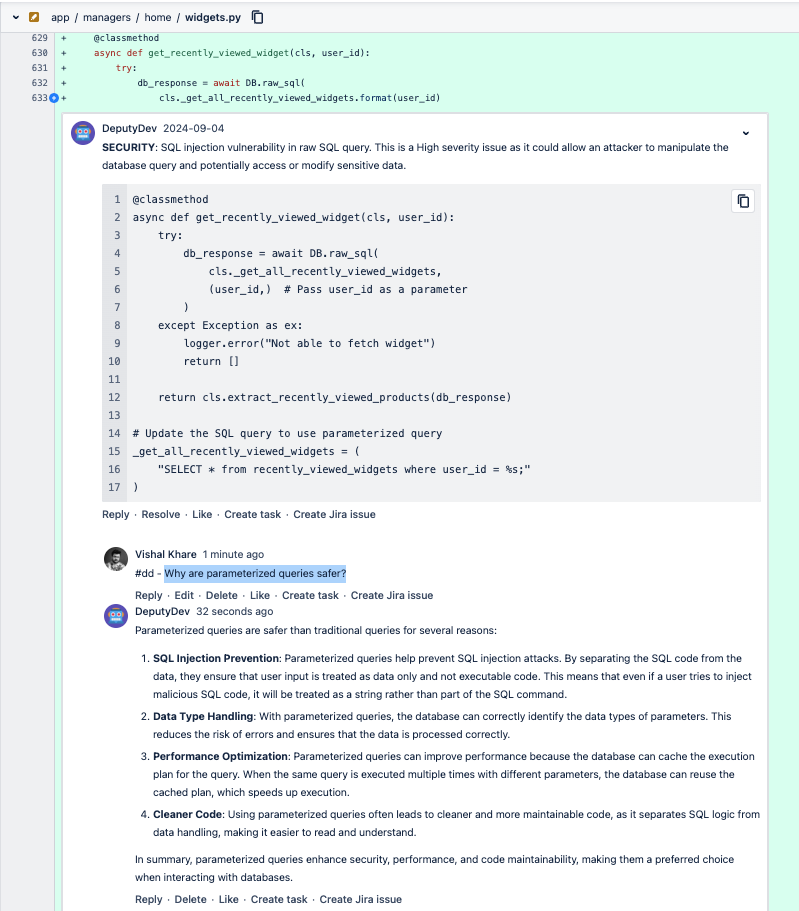}
    \caption{DeputyDev chatting can be used as a learning companion. It can help out developers and reviewers know of best practices and how to implement them.}
    \label{fig:dd_param_query}
\end{figure*}

\begin{figure*}[htbp]
    \centering
    \includegraphics[width=0.8\textwidth]
    {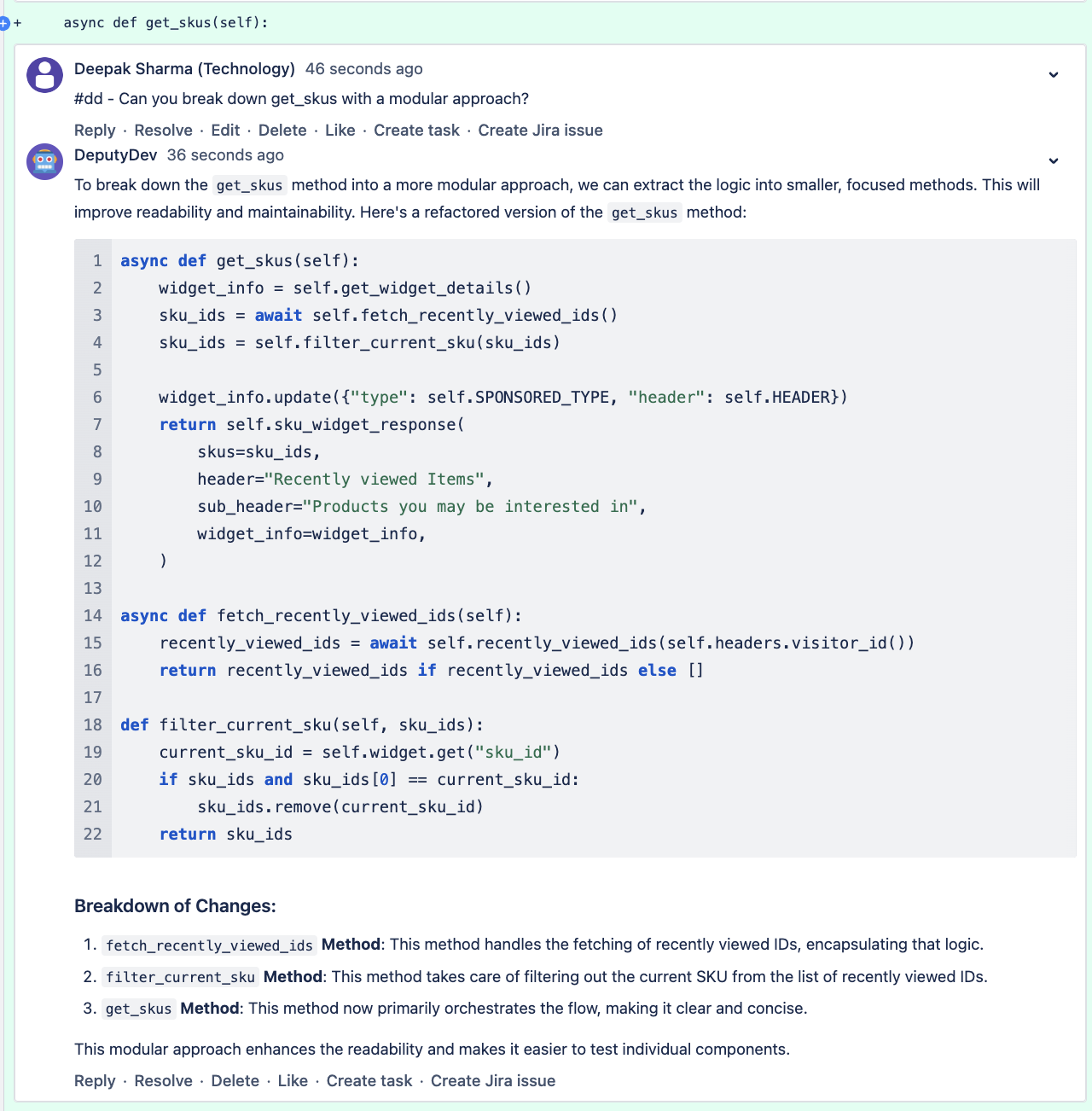}
    \caption{DeputyDev can refactor code. Converting a complex, relatively large function to modular pieces.}
    \label{fig:dd_modular_code}
\end{figure*}

\begin{figure*}[htbp]
    \centering
    \includegraphics[scale=0.40]
    {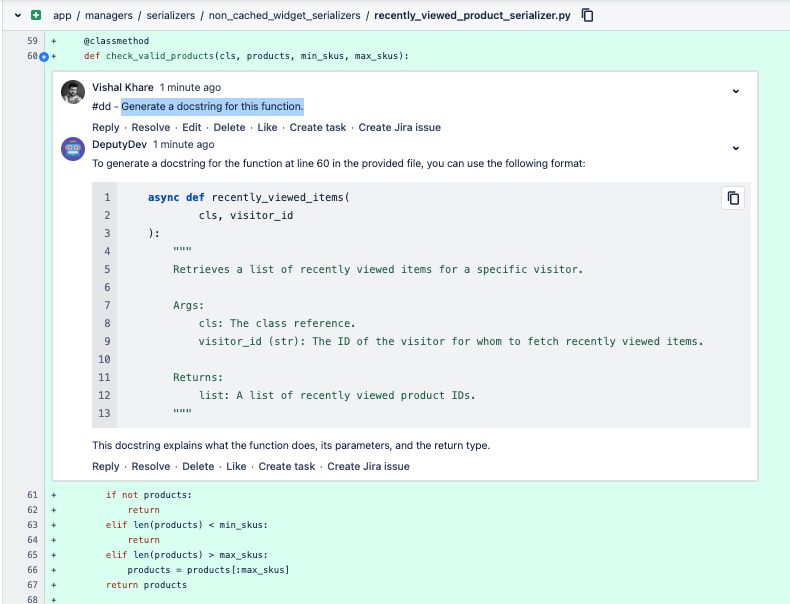}
    \caption{DeputyDev acts as a powerful tool to generate docstrings and documentations.}
    \label{fig:dd_docstring}
\end{figure*}

DeputyDev users can perform context-aware chatting with it. This is a specifically useful feature which can act as a learning companion. This feature can be used for generating code, tests, docstrings and much more. The feature is initialized by starting the comment with \texttt{\#dd} prefix. \ref{fig:dd_param_query} \ref{fig:dd_modular_code} \ref{fig:dd_docstring}

\subsection{PR summary}
\begin{figure*}[htbp]
    \centering
    \includegraphics[scale=0.50]
    {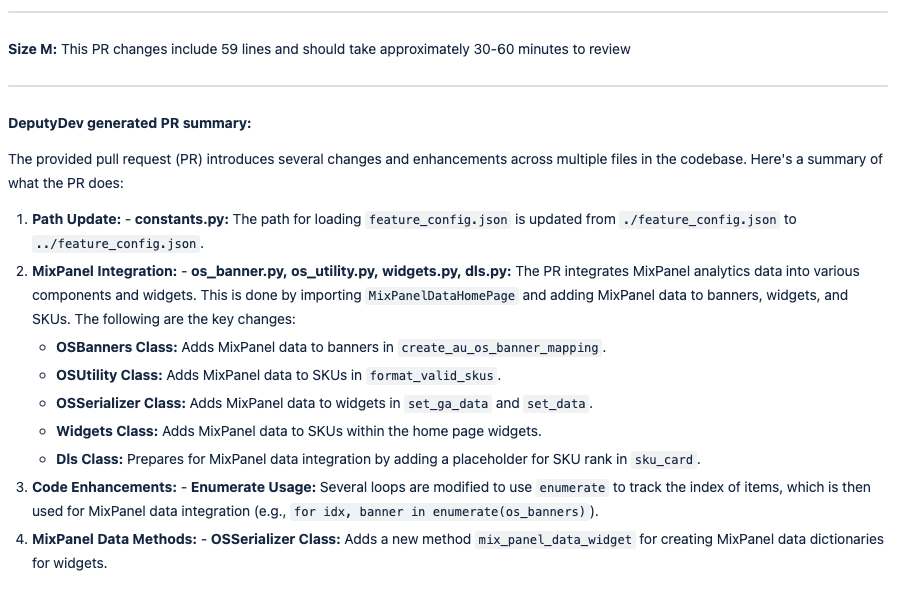}
    \caption{DeputyDev generated PR summary}
    \label{fig:dd_comm_summary}
\end{figure*}

DeputyDev generates PR summary along with sizing and review time estimates. This feature is specifically helpful for reviewers and service owners to get a gist of the change without actually going through the diff. \ref{fig:dd_comm_summary}

\section{Mathematical calculations}
\label{app:calculations}

\subsection*{1. Average code review time}

\begin{equation}
\mathlarger{\mathlarger{\text{Average code review time} = \frac{\sum_{i=1}^{n} \text{code review time}_i}{n}}}
\end{equation}

Where $n$ is the total number of pull requests.

\subsection*{2. Average code review time per LOC (Lines of Code)}

\begin{equation}
\mathlarger{\mathlarger{\text{Average code review time per LOC} = \frac{\sum_{i=1}^{n} \frac{\text{code review time}_i}{\text{lines of code}_i}}{n}}}
\end{equation}

\subsection*{3. Median code review time}

\begin{equation}
\mathlarger{\mathlarger{
\text{Median code review time} = 
\begin{cases}
    x_{\frac{n+1}{2}}, & \text{if $n$ is odd} \\[2ex]
    \frac{x_{\frac{n}{2}} + x_{\frac{n}{2}+1}}{2}, & \text{if $n$ is even}
\end{cases}
}}
\end{equation}

Where $x_1, x_2, ..., x_n$ are the sorted code review times in ascending order.
\end{appendices}

\end{document}